\documentclass{yuanbao}
\usepackage{paper-tables}
\usepackage{tikz}
\usepackage{placeins}
\usepackage{float}
\usetikzlibrary{arrows.meta,shadows}
\ifdefined\figserif\else
  \newfontfamily\figserif{STIXTwoText}[Extension=.otf,UprightFont=*-Regular,BoldFont=*-Bold,ItalicFont=*-Italic,BoldItalicFont=*-BoldItalic]
\fi

\newcommand{\vx}{\mathbf{x}}

\newcommand{\vmu}{\symbfit{\mu}}
\newcommand{\vz}{\mathbf{z}}
\newcommand{\vzhat}{\hat{\mathbf{z}}}
\newcommand{\vs}{\mathbf{s}}
\newcommand{\vy}{\mathbf{y}}
\newcommand{\vv}{\mathbf{v}}
\newcommand{\Lcal}{\mathcal{L}}
\newcommand{\model}{Tenlai-TTS~2.0}

\title{Tenlai-TTS~2.0: Continuous-Representation LLM-Based TTS System with End-to-End Training}
\shorttitle{Tenlai-TTS~2.0 Technical Report}
\reportheaderdate{2026-11-02}
\author{Changfeng Gao\texorpdfstring{\textsuperscript{*}}{}, Yong Ren\texorpdfstring{\textsuperscript{*}}{}, Jun Yuan, Zhengkun Tian, Ye Bai, Zhao You, Shidong Shang}
\affiliation{Tencent Yuanbao, Tencent Ethereal Audio Lab}
\reportteam{Tencent Yuanbao}
\date{}

\begin{document}
\maketitle
\begingroup
  \renewcommand{\thefootnote}{\fnsymbol{footnote}}
  \footnotetext[1]{Equal contribution.}
\endgroup

\begin{abstract}
Autoregressive text-to-speech (TTS) with continuous speech latents avoids quantization-induced information loss, but typically relies on a latent space that is learned separately and then frozen before language-model training.
Such a space must satisfy two competing requirements: it must retain enough acoustic detail for high-fidelity reconstruction, yet be sufficiently structured and regularized to support autoregressive prediction by a language model.
We present \model{}, which learns continuous speech representations through end-to-end training to balance acoustic detail with autoregressive predictability.
The model comprises a speech encoder with a capacity-controlled Gaussian bottleneck, a large language model (LLM) with an autoregressive diffusion transformer (AR-DiT) head, a flow-matching (FM) mel-spectrogram decoder, and a multi-task recognition model (RM).
With energy normalization and fixed-variance Gaussian noise, the bottleneck has a closed-form upper bound on its information rate.
Through joint optimization with reconstruction, recognition, and next-patch prediction objectives, we learn continuous speech representations that suit all three tasks and train the LLM and AR-DiT for autoregressive generation.
The LLM and AR-DiT are further optimized using differentiable reconstruction and recognition feedback on their predicted latents through the FM decoder and RM.
On Seed-TTS-eval, this feedback-based optimization reduces character error rate (CER) from 0.82\% to 0.74\% on test-zh and from 8.32\% to 7.08\% on test-hard-zh, and word error rate (WER) from 1.32\% to 1.27\% on test-en, compared with joint training alone.
Subsequent reinforcement learning further reduces CER to 6.29\% on test-hard-zh and WER to 1.08\% on test-en, with speaker similarity scores of 0.811, 0.766, and 0.784 on test-zh, test-en, and test-hard-zh, respectively. The demo page for \model{} is available at \url{https://tenlai-tts.github.io/Tenlai-TTS-2.0/}.
\end{abstract}
\clearpage

\section{Introduction}
\label{sec:intro}

LLM-based TTS systems now commonly synthesize speech by predicting a sequence of speech representations with an autoregressive language model and then rendering them into audio~\cite{DBLP:journals/corr/abs-2301-02111,anastassiou2024seed,du2024cosyvoice2,du2025cosyvoice3}.
Most of these systems use \emph{discrete} speech tokens.
Quantization makes the LLM's job easier through categorical next-token prediction, but the limited information budget of discrete tokens restricts how much acoustic detail they can carry, placing a ceiling on reconstruction fidelity and expressiveness.
A growing line of work therefore predicts \emph{continuous} latents directly, using a regression or diffusion head on top of the LLM to retain richer acoustic information~\cite{DBLP:journals/corr/abs-2407-08551,jia2025ditar,peng2025vibevoice,zhou2025voxcpm}.

Continuous representations do not remove the underlying tension; their larger information budget makes the balance between linguistic structure and acoustic detail more difficult.
On the one hand, the latent must contain rich acoustic information so that a decoder can reconstruct high-quality, expressive audio.
On the other hand, it must expose predictable linguistic structure so that an autoregressive model can generate it stably frame after frame.
A reconstruction-oriented autoencoder can store fine acoustic details that are hard to predict, making stable generation difficult even when reconstruction is good.
Most continuous TTS systems therefore search for a hand-tuned combination of semantic and acoustic information, for example by regularizing a VAE, distilling semantic features into the latent, or separating semantic planning from acoustic rendering, and then freeze the resulting space before the LLM is trained~\cite{lian2026dots,zhou2025voxcpm,shen2026fireredtts3,joyai2026voice}.
In this \emph{stage-wise} recipe the LLM cannot shape that balance in the representation it has to model, nor is it told what its prediction errors cost after decoding: the regression or diffusion loss does not explicitly distinguish errors that change the words or speaker from those with little audible effect.

Our previous work~\cite{gao2026e2e} addressed this separation for discrete-token TTS by training the tokenizer, the LLM, the flow-matching (FM) decoder and a recognition-based reward model (RM) end to end, and by passing straight-through Gumbel-Softmax samples from the LLM back into the FM and RM.
Joint training lets the TTS model determine which information its representation should retain for recognition, generation and reconstruction, rather than fixing the semantic--acoustic balance in advance.
This report extends that idea to continuous representations, where differentiable latents remove the need for straight-through gradient approximations through quantization and Gumbel-Softmax token sampling.
The extension still requires explicit capacity control, since a continuous latent can in principle hold an unbounded amount of information, and a differentiable feedback path from the FM and RM through the LLM's diffusion head.

We present \model{}, a continuous-representation TTS model trained end to end, and make the following contributions:
\begin{itemize}
  \item \textbf{A capacity-controlled continuous bottleneck.} Energy normalization plus fixed-variance Gaussian noise turns the latent into an additive Gaussian channel with a closed-form capacity bound: 11.1\,kbps in our main setting, about 5 times the RVQ tokenizer of Qwen3-TTS and 30--56 times the single-codebook tokenizers of CosyVoice~3.
  \item \textbf{End-to-end representation learning.} The encoder is trained together with recognition heads, the FM decoder and the LLM, so that the representation is shaped by both reconstruction and generation.
  \item \textbf{Feedback-based optimization of latent predictions.} Reconstruction and recognition objectives evaluated on predicted latents provide differentiable feedback to the LLM through the FM decoder and RM model.
  \item \textbf{Controllable decoding.} CFG strength and independent prompt routing to the LLM, AR-DiT head and FM decoder provide inference-time controls over intelligibility, speaker similarity and speech-rate following.
  \item \textbf{Results.} The feedback-based optimization stage reduces the Seed-TTS-eval CER from 0.82\% to 0.74\% on test-zh and from 8.32\% to 7.08\% on test-hard relative to the jointly trained model, at unchanged speaker similarity; a subsequent RL stage further lowers the hard-set CER to 6.29\% and the English WER to 1.08\% while raising speaker similarity on all three sets.
\end{itemize}

\section{Related Work}
\label{sec:related}

\paragraph{LLM based TTS models with discrete and continuous representation.}
Recent TTS models primarily generate speech using either discrete tokens or continuous representations.
Discrete-token systems such as Seed-TTS~\cite{anastassiou2024seed}, CosyVoice~2/3~\cite{du2024cosyvoice2,du2025cosyvoice3} and Qwen3-TTS~\cite{hu2026qwen3} model speech as sequences of quantized tokens conditioned on text and reference audio, then decode the generated tokens into speech.
Discrete tokens support categorical next-token prediction, whereas continuous representations avoid the information loss introduced by quantization.
Among continuous models, MELLE~\cite{DBLP:journals/corr/abs-2407-08551} predicts mel spectrogram frames directly, while DiTAR~\cite{jia2025ditar}, VibeVoice~\cite{peng2025vibevoice} and VoxCPM~\cite{zhou2025voxcpm} generate continuous speech latents and reconstruct audio with waveform decoders.
More recent systems, including Dots.TTS~\cite{lian2026dots}, FireRedTTS3~\cite{shen2026fireredtts3} and JoyAI-Voice~2.0~\cite{joyai2026voice}, further explore continuous representations that combine semantic structure with acoustic detail.

\paragraph{Stability and expressiveness in continuous AR TTS.}
Continuous latents preserve fine acoustic detail, but no quantization step pulls an imperfect prediction back to a valid value, so small errors can accumulate over long rollouts~\cite{lian2026dots}.
Dots.TTS~\cite{lian2026dots} extends a reconstruction-oriented VAE with multi-task recognition heads and WavLM alignment, feeds the LLM a semantic summary of each patch, conditions the flow-matching head on the full history, and post-trains that head with reward-free self-correction.
FireRedTTS3~\cite{shen2026fireredtts3} drops KL regularization to avoid over-compressing acoustic detail and instead aligns its latents with a frozen audio-understanding encoder, addressing error accumulation at the representation level.
JoyAI-Voice~2.0~\cite{joyai2026voice} splits each VAE-latent patch into a semantic representation from a frozen ASR backbone, which stabilizes planning, and a jointly trained acoustic representation, which preserves detail, and then applies supervised fine-tuning and reinforcement learning.
These works differ mainly in how they balance the stability of autoregressive generation against the expressiveness of audio reconstruction.
All three fix the latent space before generation training and reconcile the two goals afterwards, through semantic constraints, dedicated modules or post-training.
\model{} instead trains the representation and the generator end to end: the encoder receives gradients from reconstruction, recognition and generation at the same time, so the two goals reach a natural balance within the capacity set by the bottleneck.


\paragraph{Joint and feedback-based optimization.}
Several works let downstream objectives shape upstream modules: TadiCodec and DiffCodec train tokenizers with diffusion reconstruction~\cite{wang2026tadicodec,wang2026diffcodec}, JoyVoice jointly optimizes the LLM and the FM decoder~\cite{yu2025joyvoice}, and DiffRO back-propagates recognition-based rewards through a differentiable codec back end~\cite{gao2025diffro}.
Our previous work~\cite{gao2026e2e} trained the tokenizer, LLM, FM and RM of a discrete-token system end to end and fed straight-through Gumbel-Softmax samples from the LLM into the FM and RM.
\model{} carries this approach over to continuous latents, where it additionally requires explicit capacity control, and where differentiable feedback from reconstruction and recognition objectives propagates through the diffusion head's one-step latent prediction instead of through sampled tokens.

\section{Method}
\label{sec:method}

\begin{figure}[htbp]
  \centering
  \includegraphics[width=0.95\linewidth]{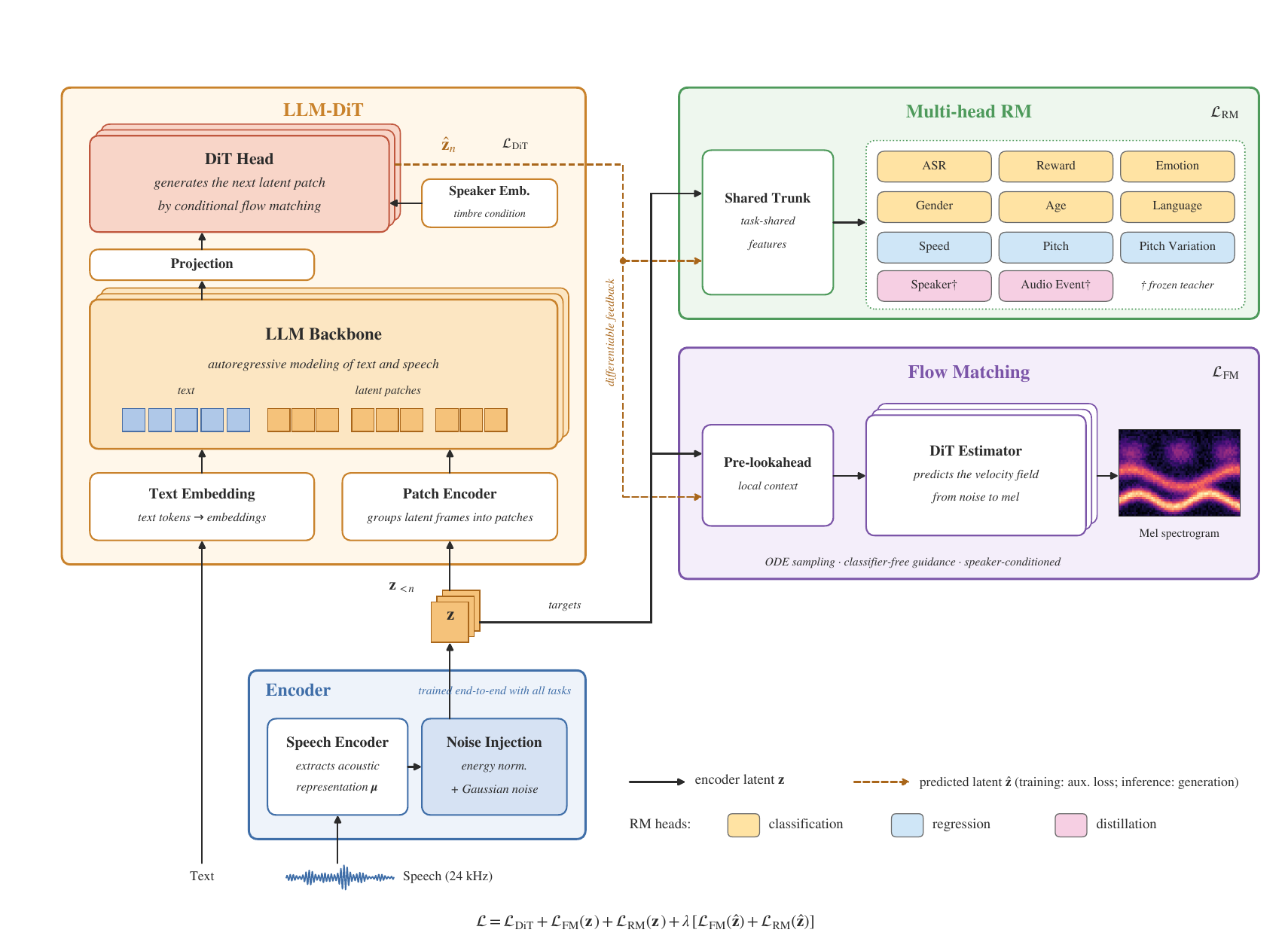}
  \caption{Overview of \model{}. Solid arrows carry the encoder latent $\vz$; dashed arrows carry the latent $\vzhat$ predicted by the LLM, which is passed to the FM decoder and the RM for feedback-based optimization during training (Sec.~\ref{sec:method:feedback}).}
  \label{fig:arch}
\end{figure}

The defining feature of \model{} is \emph{end-to-end representation learning}.
Unlike the stage-wise recipe that first trains a VAE to obtain a continuous speech representation and then trains an LLM to predict that fixed representation, \model{} learns the representation within the TTS system itself.
The speech encoder is optimized jointly with the LLM and its autoregressive diffusion head and with the flow-matching (FM) decoder, rather than trained as a separate representation model and frozen before generation training.
Consequently, the latent is shaped by both autoregressive prediction and audio reconstruction: it must be predictable by the LLM while retaining the acoustic information needed by the FM decoder.
The recognition model (RM) further constrains preservation of spoken content and speaker identity.

These objectives are balanced within an explicit information budget, imposed by energy normalization and fixed-variance Gaussian noise injection (Sec.~\ref{sec:method:encoder}).
Within this budget, joint training determines which information to retain for prediction and reconstruction, instead of asking the LLM to fit a representation chosen for reconstruction alone.
Figure~\ref{fig:arch} shows the encoder, RM (Sec.~\ref{sec:method:rm}), FM decoder and vocoder (Sec.~\ref{sec:method:decoders}), and LLM with an autoregressive diffusion head (Sec.~\ref{sec:method:lm}).
Section~\ref{sec:method:feedback} describes feedback-based optimization of latent predictions; the encoder remains trainable in both training stages (Sec.~\ref{sec:training}).

\subsection{Capacity-Controlled Continuous Bottleneck}
\label{sec:method:encoder}

A continuous latent with no information constraint can retain fine acoustic details that aid reconstruction but are difficult to predict autoregressively.
Excessive compression, in contrast, can discard information needed for high-fidelity reconstruction.
We use noise injection to constrain how much information the encoder can transmit, allowing joint training to learn a representation suited to both tasks within a controllable capacity bound.

\paragraph{Implementation.}
A speech encoder maps the input audio to a sequence of $D$-dimensional vectors $\tilde{\vmu}_t$ at 25\,Hz ($D=384$).
Before the latent is handed to any downstream module, two operations are applied:
\begin{equation}
  \begin{gathered}
  \vmu_t = \tilde{\vmu}_t \,/\, \operatorname{RMS}(\tilde{\vmu}), \\
  \vz_t = \vmu_t + \sigma\symbfit{\epsilon}_t,\quad \symbfit{\epsilon}_t\sim\mathcal{N}(\mathbf{0},\mathbf{I}).
  \end{gathered}
  \label{eq:rms}
\end{equation}
The first normalizes the latent to unit average power per dimension with a single global scale and the second injects Gaussian noise with a \emph{fixed} standard deviation $\sigma$ to limit the information available to the downstream prediction and reconstruction modules.

\paragraph{Analysis.}
For representation $z$, each dimension can be regarded as an additive Gaussian noise channel with unit input power, and by the Shannon capacity formula~\cite{shannon1948mathematical} the information per frame is bounded by
\begin{equation}
  C = \frac{D}{2}\log_2\!\left(1+\frac{1}{\sigma^2}\right)\ \text{bits}
  \label{eq:capacity}
\end{equation}
(Appendix~\ref{app:capacity}).
The capacity of the continuous latent is thus not unlimited: it is set by a single hyper-parameter and can be computed before training.

Table~\ref{tab:capacity} compares this bound with the nominal bit rates of discrete speech tokenizers, accounting for their different frame rates.
At $\sigma=0.5$, our bound is 11.1\,kbps, about five times that of the multi-codebook RVQ tokenizer of Qwen3-TTS (2.2\,kbps) and 30--56 times those of the single-codebook tokenizers (0.20--0.38\,kbps)~\cite{du2025cosyvoice3,hu2026qwen3}.
As observed in our experiments (Sec.~\ref{sec:exp}), reducing $\sigma$ improves RM recognition and FM reconstruction on encoder latents, yet degrades the LLM's latent prediction and ultimately worsens synthesized speech.
Thus, a representation that better supports recognition and reconstruction can still be harder to predict autoregressively.
The choice of $\sigma$ must therefore balance the LLM's prediction capability with the FM decoder's reconstruction capability, rather than maximize capacity or reconstruction quality alone.

\begin{table}[H]
  \centering
  \setlength{\tabcolsep}{4pt}
  \begin{lrbox}{\paperTableBox}
  \begin{tabular}{@{}llS[table-format=2.1]S[table-format=2.2]@{}}
    \toprule
    \textbf{System} & \textbf{Bottleneck} & {\tablehead{Frame rate\\(Hz)}} & {\tablehead{Rate bound\\(kbps)}} \\
    \midrule
    CosyVoice 3~\cite{du2025cosyvoice3} & FSQ, $3^8$ & 25 & 0.32 \\
    Qwen-Audio-3.0-TTS~\cite{xiang2026qwen} & FSQ, $3^{10}$ & 12.5 & 0.20 \\
    Qwen3-TTS-25Hz~\cite{hu2026qwen3} & VQ, 32,768 & 25 & 0.38 \\
    Qwen3-TTS-12Hz~\cite{hu2026qwen3} & RVQ, $16\times2{,}048$ & 12.5 & 2.2 \\
    \midrule
    Ours, $\sigma=0.5$ & Gaussian, 384-d & 25 & 11.1 \\
    Ours, $\sigma=0.2$ & Gaussian, 384-d & 25 & 22.6 \\
    \bottomrule
  \end{tabular}
  \end{lrbox}
  \captionsetup{width=\linewidth}
  \caption{Information capacity of speech representations used by LLM-based TTS systems. Values in kbps denote the theoretical information-rate upper bounds for discrete tokens and continuous representations.}
  \label{tab:capacity}
  \usebox{\paperTableBox}
\end{table}

\subsection{Recognition Model}
\label{sec:method:rm}

The recognition model (RM) checks that the latent keeps the content, the speaker and the paralinguistic properties of the input.
It consists of a Conformer backbone on top of $\vz$ and a set of task heads; the RM loss $\Lcal_{\text{RM}}(\cdot)$ is the sum of the losses of the active heads on a given latent sequence.
The base model uses only two heads: CTC-based speech recognition~\cite{graves2006connectionist} and speaker embedding prediction against a frozen CAM++ encoder~\cite{DBLP:conf/interspeech/WangZCC023}.
They constrain what matters most for zero-shot cloning, the spoken content and the speaker identity, and their labels are available for the whole pre-training corpus.
For supervised fine-tuning, the RM is extended with heads for the attributes that instructions control: emotion, language and dialect, gender, age, speech quality, speech rate, pitch, and audio events (the latter distilled from a frozen PANNs tagger~\cite{kong2020panns} when labels are missing).
With these heads, the recognition component of the feedback objective in Sec.~\ref{sec:method:feedback} checks not only \emph{what} is said and \emph{by whom}, but also \emph{how} it is said.

\subsection{FM Decoder and Vocoder}
\label{sec:method:decoders}

A flow-matching (FM) decoder following CosyVoice~3~\cite{du2025cosyvoice3} generates mel spectrograms from the latent, conditioned on the speaker embedding and the prompt, and a pre-trained HiFT vocoder~\cite{cosyvoice} converts them to waveforms.
We write $\Lcal_{\text{FM}}(\cdot)$ for its flow-matching loss with a given latent sequence as condition.
This FM--vocoder path is used for reconstruction and TTS synthesis.

\subsection{LLM with an Autoregressive DiT Head}
\label{sec:method:lm}

The LLM is initialized from Qwen3-1.7B-Base~\cite{yang2025qwen3} and reads an optional instruction, the text, and the speech latents.
Every three latent frames are merged into one input embedding, so the LLM runs at about 8.3 Hz.
At each speech step a small diffusion transformer head (AR-DiT) generates the next patch $\vz_n$ of three frames, while an auxiliary end-of-speech head predicts the termination token with loss $\Lcal_{\text{eos}}$.
The training target $\vz_n$ is a three-frame patch of $\vz$ in both stages.
The head is conditioned on the LLM hidden states $\vy_{\le n}$, the preceding latent patches $\vz_{<n}$ and the speaker embedding $\vs$, and is trained with conditional flow matching~\cite{DBLP:conf/iclr/LipmanCBNL23}: with $\vx_0\sim\mathcal{N}(\mathbf{0},\mathbf{I})$ and $\vx_t=t\,\vz_n+(1-t)\,\vx_0$,
\begin{equation}
  \Lcal_{\text{DiT}} = \mathbb{E}\,\big\|\hat{\vv}_n(\vx_t, t\mid \vy_{\le n}, \vz_{<n}, \vs) - (\vz_n-\vx_0)\big\|_2^2 .
  \label{eq:dit}
\end{equation}
Conditions are randomly dropped during training to enable classifier-free guidance~\cite{ho2022classifier}, and a few Euler steps are used at inference.
The LLM input and the history are taken from $\vz$, i.e.\ the same noisy space the model feeds back at inference.
At decoding time, the CFG scale and three independent switches control whether reference audio is provided to the LLM, the AR-DiT head and the FM decoder.

\subsection{Feedback-Based Optimization of Latent Predictions}
\label{sec:method:feedback}

The FM decoder is normally trained only on encoder latents, whereas at inference they receive latents generated by the LLM.
The LLM, in turn, is trained only with Eq.~\eqref{eq:dit}, which treats all latent dimensions alike and says nothing about how a prediction error affects the decoded speech.
We optimize the LLM's latent predictions using differentiable feedback from reconstruction and recognition objectives, evaluated through the FM decoder and a frozen RM.

For an interpolated patch $\vx_t$ and predicted velocity $\hat{\vv}_n$, the flow-matching identity gives a one-step estimate of the target patch $\vz_n$ from $\vz$,
\begin{equation}
  \vzhat_n = \vx_t + (1-t)\,\hat{\vv}_n ,
  \label{eq:onestep}
\end{equation}
which is differentiable with respect to the LLM and the AR-DiT head.
Concatenating all patches gives a predicted latent sequence $\vzhat$, on which we evaluate the feedback objective
\begin{equation}
  \Lcal_{\text{fb}} = \lambda_{\text{FM}}\,\Lcal_{\text{FM}}(\vzhat) + \lambda_{\text{RM}}\,\Lcal_{\text{RM}}(\vzhat).
  \label{eq:feedback}
\end{equation}
$\Lcal_{\text{FM}}(\vzhat)$ requires that the predicted latent can be decoded into the target mel spectrogram, and $\Lcal_{\text{RM}}(\vzhat)$ requires that the frozen recognition model still reads the correct content and speaker from it.
The gradients reach the LLM through Eq.~\eqref{eq:onestep}, so the generator learns which latent errors matter after decoding, while the FM decoder adapts to the distribution of predicted latents.
Because the predicted latent is continuous, no straight-through estimator is needed.

\section{Training}
\label{sec:training}

\subsection{E2E Training}

Our end-to-end training learns a continuous representation that is both predictable by the LLM and reconstructable by the FM decoder, within the information budget imposed by noise injection.
We use two successive stages (Table~\ref{tab:stages}): joint training first shapes the encoder latents through generation, reconstruction and recognition objectives; feedback-based optimization then adds supervision of the LLM's own latent predictions through the FM decoder and a frozen RM.

\begin{table}[H]
  \centering
  \begin{lrbox}{\paperTableBox}
  \begin{tabular}{@{}lcccc@{}}
    \toprule
    \multirow{2}{*}{\textbf{Component}} & \multicolumn{2}{c}{\textbf{E2E training}} & \multicolumn{2}{c}{\textbf{Target-speaker SFT}} \\
    \cmidrule(lr){2-3}\cmidrule(l){4-5}
    & \textbf{Stage 1} & \textbf{Stage 2} & \textbf{Stage A} & \textbf{Stage B} \\
    \midrule
    Encoder & Train & Train & \textit{Frozen} & \textit{Frozen} \\
    RM & Train & \textit{Frozen} & Train & \textit{Frozen} \\
    FM decoder & Train & Train & Train & \textit{Frozen} \\
    LLM + AR-DiT & Train & Train & \textit{Frozen} & Train \\
    \midrule
    Feedback loss $\Lcal_{\text{fb}}$ & -- & \checkmark & -- & \checkmark \\
    \bottomrule
  \end{tabular}
  \end{lrbox}
  \captionsetup{width=\linewidth}
  \caption{Training schedules for end-to-end (E2E) training and target-speaker supervised fine-tuning (SFT).}
  \label{tab:stages}
  \usebox{\paperTableBox}
\end{table}

\begin{figure}[htbp]
  \centering
  \begingroup
  \definecolor{fInk}{HTML}{2B2B2B}
  \definecolor{fGray}{HTML}{6B6B6B}
  \definecolor{fLLM}{HTML}{C9822B}
  \definecolor{fLLMfill}{HTML}{FFF7EA}
  \definecolor{fLLMbox}{HTML}{FCE6C2}
  \definecolor{fDiT}{HTML}{C0573E}
  \definecolor{fDiTfill}{HTML}{F8D5C8}
  \definecolor{fRM}{HTML}{4E9A5B}
  \definecolor{fRMfill}{HTML}{EEF7EE}
  \definecolor{fFM}{HTML}{7B55A8}
  \definecolor{fFMfill}{HTML}{F4EEFA}
  \definecolor{fEnc}{HTML}{3F6EA8}
  \definecolor{fEncfill}{HTML}{EEF3FA}
  \definecolor{fNoise}{HTML}{D6E2F3}
  \definecolor{fLat}{HTML}{F5C27A}
  \definecolor{fLatB}{HTML}{A9661A}
  \definecolor{fTok}{HTML}{AFC8E8}
  \definecolor{fTokB}{HTML}{4A6FA5}
  \definecolor{fCls}{HTML}{FFE3A3}
  \definecolor{fDist}{HTML}{F6CFE3}
  \tikzset{
    fig/.style={x=1cm, y=1cm, font=\figserif\fontsize{9}{11}\selectfont, text=fInk},
    panel/.style={rounded corners=5pt, line width=1.0pt},
    panel title/.style={font=\figserif\bfseries\fontsize{11}{13}\selectfont, inner sep=1pt},
    panel note/.style={font=\figserif\itshape\fontsize{7.5}{9}\selectfont, inner sep=1pt},
    loss label/.style={font=\fontsize{9}{11}\selectfont, text=fInk, inner sep=1pt},
    module/.style={rounded corners=3pt, line width=0.8pt, fill=white, align=center, inner sep=2pt, font=\figserif\itshape\fontsize{7.5}{9}\selectfont},
    stacked/.style={double copy shadow={shadow xshift=2.5pt, shadow yshift=2.5pt}},
    chip/.style={rounded corners=2pt, line width=0.5pt, draw=fGray, minimum height=0.42cm, inner sep=1pt, font=\figserif\fontsize{8}{9}\selectfont},
    token/.style={line width=0.4pt, minimum size=0.26cm, inner sep=0pt},
    latent/.style={line width=0.6pt, draw=fLatB, fill=fLat, minimum size=0.62cm, inner sep=0pt, rounded corners=1pt},
    flow/.style={-{Stealth[length=2.2mm, width=1.6mm]}, draw=fInk, line width=0.8pt},
    pred flow/.style={-{Stealth[length=2.2mm, width=1.6mm]}, draw=fLatB, line width=0.85pt, dash pattern=on 3.2pt off 1.4pt},
    new flow line/.style={draw=fLatB, line width=1.4pt, dash pattern=on 4pt off 1.6pt},
    new flow/.style={-{Stealth[length=2.8mm, width=2.1mm]}, new flow line},
    note/.style={font=\figserif\itshape\fontsize{7.5}{9}\selectfont, text=fInk, inner sep=1pt},
  }
  \newcommand{\boxtext}[2]{{\upshape\bfseries\fontsize{8.5}{10}\selectfont #1}\\[1pt]#2}
  \newcommand{\trainingstage}[1]{%
    \begin{tikzpicture}[fig]
      \ifnum#1=2 \def\figfade{0.28}\else \def\figfade{1}\fi
      \begin{scope}[transparency group, opacity=\figfade]
      \draw[panel, draw=fEnc, fill=fEncfill] (0.2,0.9) rectangle (5.2,3.9);
      \node[panel title, text=fEnc, anchor=west] at (0.4,3.57) {Encoder};
      \node[panel note, text=fEnc, anchor=east] at (5.05,3.57) {trained end-to-end};
      \node[module, draw=fEnc, minimum width=2.1cm, minimum height=1.3cm] (se) at (1.4,2.15) {\boxtext{Speech Encoder}{extracts acoustic\\representation $\vmu$}};
      \node[module, draw=fEnc, fill=fNoise, minimum width=2.1cm, minimum height=1.3cm] (ni) at (3.85,2.15) {\boxtext{Noise Injection}{energy norm.\\+ Gaussian noise}};
      \draw[flow] (se) -- (ni);
      \draw[fEnc, line width=0.55pt, domain=0:1.5, samples=220, smooth, variable=\t]
        plot ({0.65+\t}, {0.25 + 0.17*sin(\t*4200)*exp(-((\t-0.75)^2)/0.09)});
      \node[anchor=west, font=\figserif\fontsize{8.5}{10}\selectfont] at (2.3,0.25) {Speech (24\,kHz)};
      \draw[flow] (1.4,0.5) -- (se.south);
      \node[latent, stacked] (z) at (5.9,2.15) {$\vz$};
      \draw[flow] (ni.east) -- (z.west);
      \end{scope}
      \draw[panel, draw=fLLM, fill=fLLMfill] (7.25,0.3) rectangle (11.9,4.9);
      \node[panel title, text=fLLM] at (9.55,4.57) {LLM-DiT};
      \node[module, stacked, draw=fDiT, fill=fDiTfill, line width=0.9pt, minimum width=3.9cm, minimum height=1.15cm] (dit) at (9.55,3.4) {\boxtext{DiT Head}{generates the next latent patch\\by conditional flow matching}};
      \node[module, stacked, draw=fLLM, fill=fLLMbox, line width=0.9pt, minimum width=3.9cm, minimum height=1.6cm] (llm) at (9.55,1.5) {};
      \node[font=\figserif\bfseries\fontsize{8.5}{10}\selectfont] at (9.55,2.0) {LLM Backbone};
      \node[note] at (9.55,1.67) {autoregressive text\textendash speech modeling};
      \foreach \i in {0,1,2} \node[token, draw=fTokB, fill=fTok] at (8.2+0.3*\i,1.12) {};
      \foreach \i in {0,1,2} \node[token, draw=fLatB, fill=fLat] at (9.35+0.3*\i,1.12) {};
      \foreach \i in {0,1,2} \node[token, draw=fLatB, fill=fLat] at (10.4+0.3*\i,1.12) {};
      \draw[flow] (llm.north) -- (dit.south);
      \draw[panel, draw=fRM, fill=fRMfill] (13.0,2.75) rectangle (19.4,4.9);
      \node[panel title, text=fRM] at (16.6,4.57) {Multi-head RM};
      \ifnum#1=1
        \node[panel note, text=fRM, anchor=west] at (13.15,4.57) {trained jointly};
      \fi
      \node[module, draw=fRM, minimum width=2.2cm, minimum height=1.0cm] (trunk) at (14.4,3.45) {\boxtext{Shared Trunk}{Conformer}};
      \draw[fRM, line width=0.7pt, dash pattern=on 0.8pt off 1.3pt, rounded corners=3pt, fill=white] (15.8,2.95) rectangle (19.15,3.95);
      \node[chip, fill=fCls, minimum width=1.45cm] at (16.65,3.64) {ASR (CTC)};
      \node[chip, fill=fDist, minimum width=1.45cm] at (18.3,3.64) {Speaker$^{\dagger}$};
      \node[note, font=\figserif\itshape\fontsize{7}{8}\selectfont] at (17.47,3.18) {$^{\dagger}$frozen teacher};
      \draw[flow] (trunk.east) -- (15.8,3.45);
      \draw[panel, draw=fFM, fill=fFMfill] (13.0,0.3) rectangle (19.4,2.45);
      \node[panel title, text=fFM] at (16.2,2.12) {Flow Matching};
      \node[module, draw=fFM, minimum width=2.0cm, minimum height=1.0cm] (pre) at (14.3,1.15) {\boxtext{Pre-lookahead}{local context}};
      \node[module, stacked, draw=fFM, minimum width=2.0cm, minimum height=1.0cm] (est) at (16.5,1.15) {\boxtext{DiT Estimator}{noise $\rightarrow$ mel}};
      \node[inner sep=0pt] (mel) at (18.55,1.3) {\includegraphics[width=1.3cm,height=0.75cm]{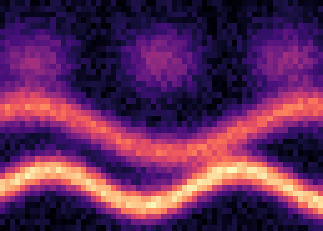}};
      \node[font=\figserif\fontsize{7.5}{9}\selectfont, anchor=north, inner sep=1pt] at (mel.south) {mel};
      \draw[flow] (pre) -- (est);
      \draw[flow] (est.east) ++(0.1,0) -- (mel.west);
      \node[font=\figserif\fontsize{8.5}{10}\selectfont] (text) at (8.5,-0.3) {Text};
      \draw[flow] (text.north) -- (8.5,0.7);
      \begin{scope}[transparency group, opacity=\figfade]
      \node[loss label, anchor=east] at (11.75,4.57) {$\Lcal_{\text{DiT}}$};
      \ifnum#1=1
        \node[loss label, anchor=east] at (19.25,4.57) {$\Lcal_{\text{RM}}$};
        \node[loss label, anchor=east] at (19.25,2.12) {$\Lcal_{\text{FM}}$};
      \fi
      \draw[fInk, line width=0.8pt] (z.east) -- (6.6,2.15);
      \draw[flow] (6.6,2.15) |- (llm.west);
      \draw[flow] (6.6,2.15) |- (dit.west);
      \node[note, anchor=east, align=right] at (6.52,1.62) {history};
      \node[note, anchor=east, align=right] at (6.52,3.66) {target\\(stop-grad)};
      \draw[fInk, line width=0.8pt] (6.6,3.4) -- (6.6,5.35) -- (12.62,5.35) -- (12.62,1.35);
      \fill[fInk] (6.6,2.15) circle (1.5pt) (6.6,3.4) circle (1.5pt) (12.62,3.65) circle (1.5pt);
      \draw[flow] (12.62,3.65) -- (13.3,3.65);
      \draw[flow] (12.62,1.35) -- (13.3,1.35);
      \ifnum#1=2
        \draw[flow] (0.3,5.05) -- (1.1,5.05);
        \node[note, anchor=west, font=\figserif\fontsize{8}{9}\selectfont] at (1.2,5.05) {encoder latent $\vz$ (Stage-1 paths, unchanged)};
      \fi
      \end{scope}
      \ifnum#1=2
        \draw[new flow line] (11.62,3.3) -- (12.3,3.3) -- (12.3,0.95);
        \draw[new flow] (12.3,3.3) -- (13.3,3.3);
        \draw[new flow] (12.3,0.95) -- (13.3,0.95);
        \fill[fLatB] (12.3,3.3) circle (2pt);
        \node[font=\fontsize{10}{12}\selectfont, text=fLatB, anchor=south, inner sep=1pt] at (11.88,3.34) {$\vzhat$};
        \node[note, text=fLatB, rotate=90, font=\figserif\bfseries\itshape\fontsize{7}{8}\selectfont] at (12.06,1.9) {differentiable feedback};
        \node[panel note, text=fRM, anchor=west, font=\figserif\bfseries\itshape\fontsize{8.5}{10}\selectfont] at (13.15,4.57) {frozen};
        \node[loss label, text=fLatB, anchor=east] at (19.25,4.57) {$\Lcal_{\text{RM}}(\vzhat)$};
        \node[loss label, text=fLatB, anchor=east] at (19.25,2.12) {$\Lcal_{\text{FM}}(\vzhat)$};
        \draw[new flow] (0.3,4.5) -- (1.1,4.5);
        \node[note, anchor=west, font=\figserif\bfseries\fontsize{8}{9}\selectfont, text=fLatB] at (1.2,4.5) {predicted latent $\vzhat$ (new in Stage~2)};
        \node[font=\fontsize{10}{12}\selectfont] at (9.8,-0.95) {${\color{fInk!35}\Lcal = \Lcal_{\text{DiT}} + \Lcal_{\text{eos}} + \Lcal_{\text{FM}}(\vz) + \Lcal_{\text{RM}}(\vz) +{}} {\color{fLatB}\lambda_{\text{FM}}\Lcal_{\text{FM}}(\vzhat) + \lambda_{\text{RM}}\Lcal_{\text{RM}}(\vzhat)}$};
      \else
        \node[font=\fontsize{10}{12}\selectfont] at (9.8,-0.95) {$\Lcal = \Lcal_{\text{DiT}} + \Lcal_{\text{eos}} + \Lcal_{\text{FM}}(\vz) + \Lcal_{\text{RM}}(\vz)$};
      \fi
    \end{tikzpicture}%
  }
  \begin{subfigure}{\linewidth}
    \centering
    \resizebox{\linewidth}{!}{\trainingstage{1}}
    \caption{Stage 1: joint training.}
    \label{fig:training_stage1}
  \end{subfigure}

  \vspace{3mm}
  \begin{subfigure}{\linewidth}
    \centering
    \resizebox{\linewidth}{!}{\trainingstage{2}}
    \caption{Stage 2: feedback-based optimization.}
    \label{fig:training_stage2}
  \end{subfigure}
  \endgroup
  \caption{Training-flow comparison. Solid arrows carry the encoder latent; thick dashed arrows highlight the predicted latent added in Stage~2, which is passed to the FM decoder and the frozen RM to provide reconstruction and recognition feedback (brown terms)}
  \label{fig:training_stages}
\end{figure}

\paragraph{Stage 1: joint training.}
The encoder, RM, FM decoder, LLM and AR-DiT head are trained jointly with reconstruction and recognition losses, $\Lcal_{\text{FM}}(\vz)+\Lcal_{\text{RM}}(\vz)$, together with $\Lcal_{\text{DiT}}$ and the auxiliary termination-token loss $\Lcal_{\text{eos}}$.
The encoder receives gradients from reconstruction, recognition and generation at the same time, so the representation is chosen by all tasks together rather than fixed in advance.
The AR-DiT target is detached from the encoder, which together with the energy normalization prevents the latent from collapsing.
In our runs the tasks did not conflict and appeared complementary. 

\paragraph{Stage 2: feedback-based optimization.}
The encoder continues to train jointly with the FM decoder, LLM and AR-DiT head, while the RM is frozen as a recognition reference.
The feedback objective in Eq.~\eqref{eq:feedback} is added so that predicted latents are also supervised through reconstruction and recognition.
Freezing the RM is essential: otherwise $\Lcal_{\text{RM}}(\vzhat)$ could be reduced by adapting the recognizer to the generator's errors instead of correcting them.

\subsection{Target Speaker Fine-Tuning}
\label{sec:training:speaker}

The ability to train or freeze each module independently supports a two-stage adaptation procedure (Table~\ref{tab:stages}) for target speaker fine-tuning, with the speech encoder frozen throughout to preserve the learned latent representation.

In Stage A, we adapt the FM decoder and RM on encoder latents using reconstruction and recognition losses, while keeping the LLM and AR-DiT fixed.
In Stage B, we freeze the adapted FM decoder and RM and fine-tune the LLM, latent adapters and AR-DiT with next-patch prediction and termination losses, together with the feedback objective $\Lcal_{\text{fb}}$.
The frozen acoustic and recognition modules provide differentiable supervision on the predicted latents, guiding generation toward the target speaker's acoustic characteristics while preserving linguistic content.

\subsection{Reinforcement Learning}
\label{sec:training:rl}

We use reinforcement learning to optimize content accuracy, speaker fidelity and generation stability.
We adapt Diffusion Negative-aware FineTuning (DiffusionNFT)~\cite{zheng2025diffusionnft} to autoregressive speech generation, updating the LLM, latent input adapter and AR-DiT while keeping the speech encoder, acoustic decoder and reward models fixed.

For each text-and-speaker condition $c$, a behavior policy generates $K$ latent sequences $\{Z_i\}_{i=1}^{K}$, which are decoded and assigned utterance-level rewards.
Rewards are standardized within each group and mapped to weights $u_i\in[0,1]$, with larger weights assigned to higher-reward samples.
For each generated patch $\vz_{i,n}$, we sample noise $\symbfit{\epsilon}$ and a flow time $t$ to form $\vx_t=t\vz_{i,n}+(1-t)\symbfit{\epsilon}$.
Conditioned on $c$ and the generated prefix, the current and behavior velocity predictors define the positive and negative reconstruction branches:
\begin{equation}
  \vv^{+}_{\theta}=\vv_{\theta},\qquad
  \vv^{-}_{\theta}=2\vv_b-\vv_{\theta},\qquad
  \hat{\vz}^{\,\pm}_{i,n}=\vx_t+(1-t)\vv^{\pm}_{\theta},
  \label{eq:nft_branches}
\end{equation}
where $\vv_b$ is fixed during each update.
The reward-weighted objective, up to a constant scale, is
\begin{equation}
  \Lcal_{\mathrm{NFT}}
  =\mathbb{E}\!\left[
    \frac{1}{K}\sum_{i=1}^{K}\frac{1}{N_i}\sum_{n=1}^{N_i}
    \left\{u_i\,d(\hat{\vz}^{\,+}_{i,n},\vz_{i,n})
    +(1-u_i)\,d(\hat{\vz}^{\,-}_{i,n},\vz_{i,n})\right\}
  \right],
  \label{eq:nft_speech}
\end{equation}
where $N_i$ is the number of patches and $d$ is squared reconstruction error normalized by a detached mean absolute error.
The two branches encourage higher-reward generations and suppress lower-reward ones, while averaging over patches gives each utterance equal weight regardless of length.
This forward-process update uses generated histories without requiring explicit likelihoods or gradients through rollouts and waveform rewards.
We maintain a moving-average behavior policy and regularize the learner toward its initialization to stabilize optimization.

For target-speaker adaptation, rewards combine content accuracy, similarity to the target speaker, and speaking-rate and duration consistency with reference recordings, together with penalties for truncated generations.
For zero-shot TTS, rewards combine content accuracy and utterance completion with similarity to the speaker prompt, speaking-rate consistency, and penalties for abnormal duration, silence and clipping.

\section{Experiments}
\label{sec:exp}

\subsection{Training Data}
\label{sec:training:data}

\paragraph{Data distribution.}
The general-purpose speech corpus comprises 1.6 million hours of Chinese and English speech, with Chinese dialect and regional varieties including Cantonese, Teochew, Shanghainese, Sichuan, Northeastern and Henan speech.
Source domains include educational speech, audiobooks, podcasts, film and animation, and meeting speech; whispering is also represented.
To further raise the performance ceiling of the TTS model, we additionally collected 50,000 hours of challenging samples, covering extreme speech rates, long-form speech, diverse domains, dialects and speaking styles, and child and older voices (Table~\ref{tab:training_data}).

\begin{table}[H]
  \centering
  \begin{lrbox}{\paperTableBox}
  \begin{tabularx}{0.88\linewidth}{@{}>{\RaggedRight\arraybackslash}p{0.28\linewidth}>{\RaggedRight\arraybackslash}X@{}}
    \toprule
    \textbf{Component} & \textbf{Coverage and selection} \\
    \midrule
    Quality-screened speech & Speech selected by objective whole-utterance and local-window quality estimates. \\
    \addlinespace[4pt]
    Extreme speech rates & Fast and slow speech, with targeted reinforcement of speech-rate coverage. \\
    \addlinespace[4pt]
    Long-form speech & Single-speaker examples formed by joining adjacent VAD segments from source recordings. \\
    \addlinespace[4pt]
    Broad-coverage supplement & Speech from additional domains, dialects and speaking styles. \\
    \addlinespace[4pt]
    Child and older voices & Dedicated sources spanning ASR, meeting, storytelling and audiovisual speech. \\
    \bottomrule
  \end{tabularx}
  \end{lrbox}
  \captionsetup{width=\linewidth}
  \caption{Main components of the additional 50,000-hour collection of challenging samples for improving TTS performance.}
  \label{tab:training_data}
  \usebox{\paperTableBox}
\end{table}

\paragraph{Data filtering.}
To support reliable TTS training with this diverse data collection, we combine quality-based selection with waveform- and alignment-based filtering.
For the quality-screened portion of the targeted supplement, speech is preselected using objective quality estimates at both the whole-utterance and local-window levels.
During training, we discard examples explicitly labeled as multi-speaker and waveforms containing non-finite or near-full-scale samples; examples with missing or unknown speaker-count labels are retained.
Where valid alignments are available, we trim boundary silence while preserving up to 150\,ms before speech and 200\,ms after it, and retain segments with durations of 0.05--199.8\,s after trimming.
To limit abrupt loudness changes, we reject examples whose level range or head-to-tail drift exceeds 15\,dB, measured over energy-selected voiced frames when sufficient voiced content is available.
Finally, the retained audio is taken from the first channel and resampled to 24\,kHz, with a corresponding 16\,kHz version for components operating at that sampling rate.

\subsection{Evaluation Protocol}
\label{sec:exp:protocol}

We evaluate zero-shot voice cloning on the zh, en and hard sets of Seed-TTS-eval~\cite{anastassiou2024seed}.
Chinese CER uses Paraformer~\cite{gao2023paraformerfastaccurateparallel} and WER; English WER uses Whisper-large-v3~\cite{radford2023robust}, applying Whisper's English text normalizer to both texts.
SIM is the mean whole-utterance cosine similarity to the reference prompt.

To verify the model's effectiveness on highly expressive audio, we also use 198 utterance from CV3-Eval~\cite{du2025cosyvoice3} subject test set for some experiments.

\subsection{Representation Reconstruction and Recognition}
\label{sec:exp:codec}

We evaluate encoder latents on reconstruction and recognition tasks.
FM reconstruction uses $\vz$ and source-speaker embeddings on Seed-TTS zh/en and CV3-Subject split.
Direct RM recognition uses the full CommonVoice zh/en test sets for comparison with discrete FSQ representations.

\begin{table}[H]
  \centering
  \setlength{\tabcolsep}{3.5pt}
  \begin{lrbox}{\paperTableBox}
  \begin{tabular}{@{}lS[table-format=1.2]S[table-format=2.2]*{2}{S[table-format=1.2]S[table-format=1.3]}S[table-format=2.2]S[table-format=1.3]@{}}
    \toprule
    & \multicolumn{2}{c}{\textbf{RM recognition}} & \multicolumn{6}{c}{\textbf{FM reconstruction}} \\
    \cmidrule(lr){2-3}\cmidrule(l){4-9}
    & {\textbf{CMV-zh}} & {\textbf{CMV-en}} & \multicolumn{2}{c}{\textbf{Seed-TTS zh}} & \multicolumn{2}{c}{\textbf{Seed-TTS en}} & \multicolumn{2}{c}{\textbf{CV3-Subject}} \\
    \cmidrule(lr){2-3}\cmidrule(lr){4-5}\cmidrule(lr){6-7}\cmidrule(l){8-9}
    \textbf{Representation} & {WER$\downarrow$} & {WER$\downarrow$} & {WER$\downarrow$} & {SIM$\uparrow$} & {WER$\downarrow$} & {SIM$\uparrow$} & {WER$\downarrow$} & {SIM$\uparrow$} \\
    \midrule
    FSQ $3^8$ & 7.16 & 14.77 & 3.05 & 0.825 & 3.26 & 0.704 & 11.67 & 0.781 \\
    FSQ $3^{10}$ & \bfseries 6.50 & 14.53 & 2.93 & 0.826 & \bfseries 3.08 & 0.702 & 11.56 & 0.780 \\
    \midrule
    $\sigma=0.5$ & 6.57 & 14.65 & 1.78 & 0.939 & 4.07 & 0.897 & 7.70 & 0.909 \\
    $\sigma=0.2$ & 6.55 & \bfseries 14.22 & \bfseries 1.76 & \bfseries 0.952 & 4.03 & \bfseries 0.923 & \bfseries 7.38 & \bfseries 0.920 \\
    \bottomrule
  \end{tabular}
  \end{lrbox}
  \captionsetup{width=\linewidth}
  \caption{RM recognition and FM reconstruction results. WER is reported in percent. Bold numbers indicate the best result in each column.}
  \label{tab:codec_reconstruction}
  \usebox{\paperTableBox}
\end{table}

Table~\ref{tab:codec_reconstruction} suggests that RM recognition and FM reconstruction benefit differently from a larger information budget.
Reducing $\sigma$ relaxes the capacity constraint, but yields only limited recognition gains.
This suggests that the tighter bottleneck already retains most of the linguistic information needed by the RM.
Within the continuous bottleneck, reducing $\sigma$ improves FM reconstruction speaker similarity and slightly lowers reconstruction WER across all three test sets.
Relative to FSQ, however, the continuous representations improve speaker similarity consistently and reduce reconstruction WER on Chinese and CV3; while English reconstruction WER is higher.
The additional capacity therefore appears to preserve acoustic and speaker-specific detail more than it improves linguistic content recovery.
The broadly comparable RM performance but higher reconstruction similarity relative to FSQ is consistent with this interpretation.
\FloatBarrier

\subsection{Zero-Shot TTS on Seed-TTS-eval}
\label{sec:exp:seedtts}

\begin{table}[!t]
  \centering
  \setlength{\tabcolsep}{3pt}
  \begin{lrbox}{\paperTableBox}
  \begin{tabular}{@{}lS[table-format=1.2]S[table-format=1.3]S[table-format=1.2]S[table-format=1.3]@{}}
    \toprule
    \multirow{2}{*}{\textbf{Model}} & \multicolumn{2}{c}{\textbf{test-zh}} & \multicolumn{2}{c}{\textbf{test-en}} \\
    \cmidrule(lr){2-3}\cmidrule(l){4-5}
    & {CER (\%)$\downarrow$} & {SIM$\uparrow$} & {WER (\%)$\downarrow$} & {SIM$\uparrow$} \\
    \midrule
    Seed-TTS~\cite{anastassiou2024seed} & 1.12 & 0.796 & 2.25 & 0.762 \\
    MaskGCT~\cite{wang2025maskgct} & 2.27 & 0.774 & 2.62 & 0.714 \\
    F5-TTS~\cite{DBLP:journals/corr/abs-2410-06885} & 1.53 & 0.760 & 2.00 & 0.670 \\
    DiTAR~\cite{jia2025ditar} & 1.02 & 0.753 & 1.69 & 0.735 \\
    MegaTTS3~\cite{jiang2025megatts3} & 1.52 & 0.790 & 2.79 & 0.771 \\
    MiniMax-Speech~\cite{zhang2025minimax} & 0.83 & 0.783 & 1.65 & 0.692 \\
    CosyVoice3-1.5B~\cite{du2025cosyvoice3} & 1.12 & 0.781 & 2.21 & 0.720 \\
    IndexTTS2~\cite{zhou2025indextts2} & 1.03 & 0.765 & 2.23 & 0.706 \\
    OpenAudio-s1-mini~\cite{liao2026fish} & 1.18 & 0.685 & 1.94 & 0.550 \\
    VibeVoice~\cite{peng2025vibevoice} & 1.16 & 0.744 & 3.04 & 0.689 \\
    Qwen3-TTS-1.7B-Base~\cite{hu2026qwen3} & 1.22 & 0.770 & 1.23 & 0.717 \\
    MOSS-TTS~\cite{gong2026moss} & 1.46 & 0.762 & 1.92 & 0.693 \\
    VoxCPM2~\cite{zhou2026voxcpm2} & 0.97 & 0.795 & 1.84 & 0.753 \\
    Dots.TTS~\cite{lian2026dots} & 0.94 & 0.810 & 1.30 & 0.771 \\
    Qwen-Audio-3.0-TTS~\cite{xiang2026qwen} & 0.84 & 0.792 & 1.54 & 0.762 \\
    FireRedTTS3-Base~\cite{shen2026fireredtts3} & 1.01 & 0.809 & 1.64 & \bfseries 0.772 \\
    JoyAI-Voice2.0-Base~\cite{joyai2026voice} & \bfseries 0.74 & 0.771 & 1.27 & 0.739 \\
    FishAudio-Drama3~\cite{han2026fish} & 0.83 & {--} & 1.54 & {--} \\
    \midrule
    \multicolumn{5}{@{}l}{\textbf{\model{}}} \\
    \addlinespace[1pt]
    \quad Stage 1, $\sigma=0.2$ & 0.92 & 0.765 & 1.63 & 0.697 \\
    \quad Stage 1, $\sigma=0.5$ & 0.82 & 0.787 & 1.32 & 0.745 \\
    \quad Stage 2, $\sigma=0.5$ & \bfseries 0.74 & 0.782 & 1.27 & 0.744 \\
    \quad RL, $\sigma=0.5$ & 0.77 & \bfseries 0.811 & \bfseries 1.08 & 0.766 \\
    \bottomrule
  \end{tabular}
  \end{lrbox}
  \captionsetup{width=\linewidth}
  \caption{Zero-shot Seed-TTS-eval results. Baselines are author-reported. Our final model is obtained by applying RL to the Stage~2 checkpoint. Bold numbers indicate the best result in each column, including ties.}
  \label{tab:seedtts}
  \usebox{\paperTableBox}
\end{table}

Table~\ref{tab:seedtts} shows that increasing the representation capacity does not necessarily improve zero-shot TTS.
In Stage~1, reducing $\sigma$ from 0.5 to 0.2 raises the capacity bound from 11.1 to 22.6\,kbps (Table~\ref{tab:capacity}), yet zh CER increases from 0.82\% to 0.92\% and en WER from 1.32\% to 1.63\%.
Speaker similarity also drops from 0.787 to 0.765 on zh and from 0.745 to 0.697 on en.
The degradation in both intelligibility and speaker similarity is consistent with the capacity--predictability trade-off discussed in Sec.~\ref{sec:method:encoder}: a richer encoder representation can support recognition and reconstruction while being harder for the LLM to predict.
Thus, the information budget should be selected jointly for autoregressive predictability and FM reconstruction, rather than for reconstruction capacity alone; we use $\sigma=0.5$ as the default.

At $\sigma=0.5$, Stage~2 reduces zh CER/en WER from 0.82\%/1.32\% to 0.74\%/1.27\%, with small SIM decreases from 0.787/0.745 to 0.782/0.744.
Our final model applies 1,000 RL updates to the Stage~2 checkpoint, reaching 0.77\% zh CER and 1.08\% en WER, with speaker similarity of 0.811 and 0.766, respectively.
Among the systems listed in Table~\ref{tab:seedtts}, it achieves the lowest English WER and highest Chinese speaker similarity; the effects of each training stage are examined in Sec.~\ref{sec:exp:rl}.

\subsection{Zero-Shot Acoustic-Detail Following}
\label{sec:exp:rate_comparison}

As TTS quality improves, WER and SIM alone become less discriminative of differences in voice cloning.
Subjective listening tests can reveal these differences, but require substantial evaluation effort and are sensitive to raters and test conditions, limiting reproducibility.
We therefore evaluate finer-grained acoustic-detail cloning through the correlation of synthesized speech with its reference prompt in speech rate, pitch and volume.
Speech rate characterizes how fast speech is delivered; we measure it as the number of text units per second after removing boundary silence.
Pitch is the median voiced $F_0$ estimated by pYIN; volume is the median active-frame RMS level in dBFS, without gain normalization.
We compute Pearson $r$ across valid reference--output pairs using speech rate, log $F_0$ and dBFS, respectively; higher correlation indicates stronger reference following.

For models, we compare single-codebook discrete CosyVoice3-0.5B~\cite{du2025cosyvoice3}, multi-codebook discrete FishAudio-Drama3~\cite{han2026fish} , and continuous-representation Dots.TTS~\cite{lian2026dots}, FireRedTTS3~\cite{shen2026fireredtts3} and \model{}.
The test sets are Seed-TTS zh, Seed-TTS en and CV3-Subject.

\begin{table}[H]
  \centering
  \begin{lrbox}{\paperTableBox}
  \begin{tabular}{@{}ll*{3}{S[table-format=1.3]}@{}}
    \toprule
    \multirow{2}{*}{\textbf{Test set}} & \multirow{2}{*}{\textbf{Model}} & \multicolumn{3}{c}{\textbf{Pearson $r$}$\uparrow$} \\
    \cmidrule(l){3-5}
    & & {Speech rate} & {Pitch ($\log F_0$)} & {Volume (dBFS)} \\
    \midrule
    \multirow{5}{*}{Seed-TTS zh} & FishAudio-Drama3 & 0.443 & 0.928 & 0.818 \\
                & CosyVoice3-0.5B & 0.599 & 0.966 & 0.908 \\
                & Dots.TTS & 0.633 & 0.961 & 0.918 \\
                & FireRedTTS3 & 0.669 & 0.968 & \bfseries 0.930 \\
                & \model{} & \bfseries 0.760 & \bfseries 0.973 & 0.919 \\
    \midrule
    \multirow{5}{*}{Seed-TTS en} & FishAudio-Drama3 & 0.371 & 0.910 & 0.813 \\
                & CosyVoice3-0.5B & 0.332 & 0.933 & 0.859 \\
                & Dots.TTS & \bfseries 0.425 & 0.924 & 0.890 \\
                & FireRedTTS3 & 0.410 & \bfseries 0.944 & 0.925 \\
                & \model{} & 0.401 & 0.942 & \bfseries 0.927 \\
    \midrule
    \multirow{5}{*}{CV3-sub} & FishAudio-Drama3 & 0.730 & 0.766 & 0.836 \\
                & CosyVoice3-0.5B & 0.790 & 0.913 & 0.884 \\
                & Dots.TTS & \bfseries 0.868 & 0.927 & 0.901 \\
                & FireRedTTS3 & 0.837 & 0.918 & 0.934 \\
                & \model{} & 0.850 & \bfseries 0.961 & \bfseries 0.948 \\
    \bottomrule
  \end{tabular}
  \end{lrbox}
  \captionsetup{width=\linewidth}
  \caption{Zero-shot speech rate, pitch and volume following (Pearson $r$; higher is better). Bold numbers indicate the best result for each test set and metric.}
  \label{tab:rate_models}
  \label{tab:prosody_models}
  \usebox{\paperTableBox}
\end{table}

\model{} is competitive across all three dimensions: it achieves the highest zh speech-rate correlation (0.760), the highest en volume correlation (0.927), and the highest CV3 pitch/volume correlations (0.961/0.948).
Across all three test sets, the continuous models have higher speech-rate and volume correlations than both discrete baselines.
Compared with the speech rate and volume, pitch differences are smaller. This pattern is consistent with a limited discrete information budget retaining pitch-related cues more readily than timing and volume detail, while continuous representations can preserve a broader range of acoustic information.
It suggests a higher ceiling for fine-grained voice cloning with continuous representations, beyond differences visible in WER or SIM.

\FloatBarrier

\subsection{Effects of Feedback-Based Optimization and Reinforcement Learning}
\label{sec:exp:feedback}
\label{sec:exp:rl}

Table~\ref{tab:stages_result} reports Seed-TTS-eval results after joint training (Stage~1), feedback-based optimization (Stage~2), and reinforcement learning on a curated subset of high-quality training data.

\begin{table}[H]
  \centering
  \begin{lrbox}{\paperTableBox}
  \begin{tabular}{@{}l*{3}{S[table-format=1.2]S[table-format=1.3]}@{}}
    \toprule
    \multirow{2}{*}{\textbf{Training condition}} & \multicolumn{2}{c}{\textbf{test-zh}} & \multicolumn{2}{c}{\textbf{test-en}} & \multicolumn{2}{c}{\textbf{test-hard}} \\
    \cmidrule(lr){2-3}\cmidrule(lr){4-5}\cmidrule(l){6-7}
    & {CER (\%)$\downarrow$} & {SIM$\uparrow$} & {WER (\%)$\downarrow$} & {SIM$\uparrow$} & {CER (\%)$\downarrow$} & {SIM$\uparrow$} \\
    \midrule
    Stage 1 (joint) & 0.82 & 0.787 & 1.32 & 0.745 & 8.32 & 0.757 \\
    Stage 2 & \bfseries 0.74 & 0.782 & 1.27 & 0.744 & 7.08 & 0.752 \\
    RL & 0.77 & \bfseries 0.811 & \bfseries 1.08 & \bfseries 0.766 & \bfseries 6.29 & \bfseries 0.784 \\
    \bottomrule
  \end{tabular}
  \end{lrbox}
  \captionsetup{width=\linewidth}
  \caption{Seed-TTS-eval results across successive training stages. RL starts from the Stage~2 checkpoint. Bold numbers indicate the best result in each column.}
  \label{tab:stages_result}
  \label{tab:zeroshot_rl}
  \usebox{\paperTableBox}
\end{table}

\begin{figure}[t]
  \centering
  \includegraphics[width=\linewidth]{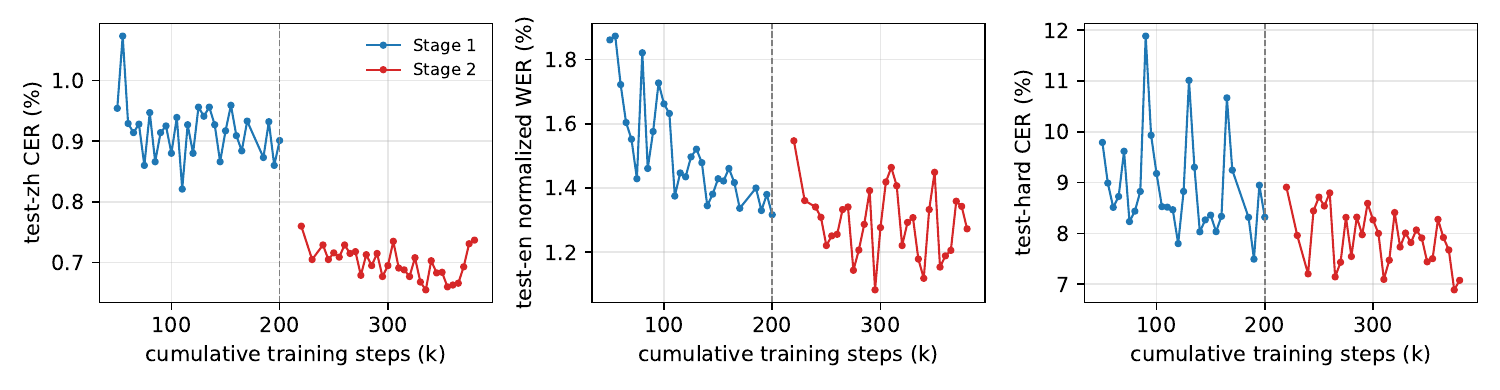}
  \caption{Seed-TTS-eval training curves: Stage~1 (blue, from 50k) and Stage~2 (red, offset by 200k). The dashed line marks the stage transition.}
  \label{fig:curve}
\end{figure}

Stage~2 lowers recognition errors on all three test sets, with small decreases in speaker similarity.
The training curves in Figure~\ref{fig:curve} show continued improvements after the Stage~1 plateau, consistent with the benefit of reconstruction and recognition feedback on latent predictions.

RL further reduces English WER from 1.27\% to 1.08\% and hard-set CER from 7.08\% to 6.29\%, corresponding to relative reductions of 15.0\% and 11.2\%.
Speaker similarity increases on all three sets, from 0.782/0.744/0.752 to 0.811/0.766/0.784 on zh/en/hard, while zh CER rises slightly from 0.74\% to 0.77\%.
These results indicate complementary benefits from feedback-based optimization and utterance-level reward optimization.

\Needspace{0.55\textheight}
\subsection{Target Speaker Fine-Tuning}
\label{sec:exp:speaker}

\begin{figure}[H]
  \centering
  \includegraphics[width=\linewidth]{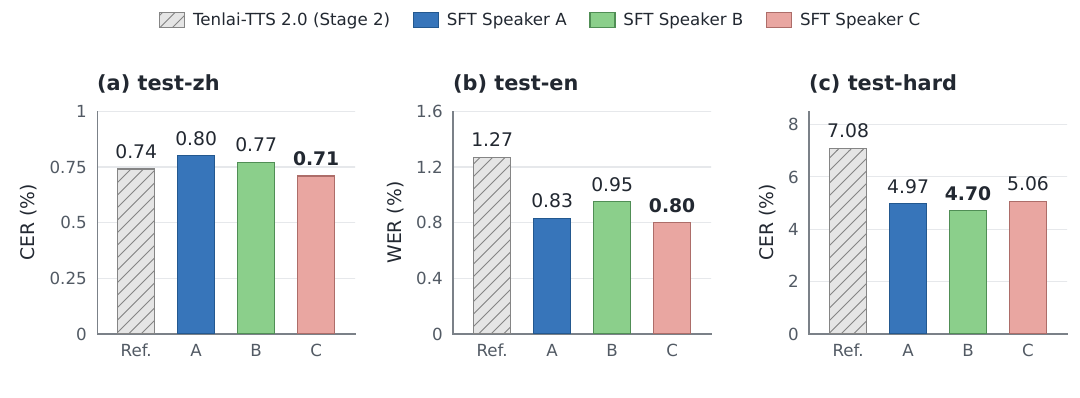}
  \caption{Content consistency on Seed-TTS-eval for three anonymized target-speaker models, compared with the \model{} Stage~2 zero-shot reference. Lower CER/WER is better; panels use different vertical scales.}
  \label{fig:target_speaker_results}
\end{figure}

We evaluate the two-stage fine-tuning procedure in Sec.~\ref{sec:training:speaker} on three anonymized target speakers using the Seed-TTS-eval texts.
Figure~\ref{fig:target_speaker_results} compares their content consistency with the Stage~2 zero-shot model reported in Table~\ref{tab:seedtts}.
The adapted models achieve Chinese CER of 0.71--0.80\%, close to the 0.74\% reference, while English WER falls to 0.80--0.95\% from 1.27\% and hard-set CER to 4.70--5.06\% from 7.08\%.
These results demonstrate that target-speaker adaptation maintains intelligibility across languages, with lower recognition errors on English and challenging Chinese texts than the zero-shot reference.

\subsection{Inference-Time Control: CFG and Prompt Routing}
\label{sec:exp:decoding}

We evaluate the effects of classifier-free guidance (CFG) and reference-audio routing on intelligibility and acoustic reference.
Routing is specified by three binary indicators for reference-audio conditioning of the LLM, AR-DiT head and FM decoder, respectively; 1 enables conditioning and 0 disables it.
For example, route 111 supplies reference audio to all three modules.
The two factors are evaluated separately: CFG is varied with routing fixed to 111, and routing is varied at CFG 1.0.
In Table~\ref{tab:cfg_cv3}, $r$ is the reference--output log-speech-rate correlation reported in the Speech rate column of Table~\ref{tab:rate_models}.

\begin{table}[H]
  \centering
  \begin{lrbox}{\paperTableBox}
  \begin{minipage}[t]{0.42\linewidth}
    \centering
    \textbf{(a) CFG sweep (route 111)}\par\vspace{4pt}
    \begin{tabular}{@{}S[table-format=1.1]S[table-format=2.2]S[table-format=1.4]S[table-format=1.3]@{}}
      \toprule
      {\textbf{CFG}} & {WER$\downarrow$} & {SIM$\uparrow$} & {$r\uparrow$} \\
      \midrule
      0.0 & 11.55 & 0.7526 & 0.864 \\
      0.1 & 9.23 & 0.7544 & 0.858 \\
      0.5 & 7.07 & \bfseries 0.7644 & \bfseries 0.866 \\
      0.7 & 6.70 & 0.7621 & 0.836 \\
      1.0 & 6.01 & 0.7581 & 0.832 \\
      1.5 & 5.61 & 0.7615 & 0.789 \\
      2.0 & \bfseries 5.46 & 0.7522 & 0.715 \\
      \bottomrule
    \end{tabular}
  \end{minipage}\hspace{0.05\linewidth}%
  \begin{minipage}[t]{0.40\linewidth}
    \centering
    \textbf{(b) Prompt routing (CFG 1.0)}\par\vspace{4pt}
    \begin{tabular}{@{}cS[table-format=1.2]S[table-format=1.4]S[table-format=1.3]@{}}
      \toprule
      \textbf{Route} & {WER$\downarrow$} & {SIM$\uparrow$} & {$r\uparrow$} \\
      \midrule
      \texttt{111} & 6.01 & 0.7581 & \bfseries 0.832 \\
      \texttt{011} & 3.59 & \bfseries 0.7661 & 0.829 \\
      \texttt{001} & \bfseries 2.99 & 0.7555 & 0.827 \\
      \bottomrule
    \end{tabular}
  \end{minipage}
  \end{lrbox}
  \captionsetup{width=\linewidth}
  \caption{Inference-time CFG and prompt-routing control. CFG is varied with route 111; routing is varied at CFG 1.0. Bold numbers indicate the best result within each panel. WER is reported in percent.}
  \label{tab:cfg_cv3}
  \label{tab:prompt_cv3}
  \usebox{\paperTableBox}
\end{table}

\paragraph{CFG for AR-DiT.}
With route 111, increasing CFG from 0 to 2.0 reduces WER from 11.55\% to 5.46\% (52.7\% relative), while speaker similarity and speech-rate correlation peak at CFG 0.5 (0.7644 and 0.866, respectively; Table~\ref{tab:cfg_cv3}).
Speech-rate correlation falls from 0.866 at CFG 0.5 to 0.715 at 2.0, indicating that stronger guidance favors intelligibility over acoustic reference following.

\paragraph{Prompt routing.}
Removing the LLM prompt (011) lowers WER from 6.01\% to 3.59\% and increases SIM from 0.7581 to 0.7661; FM-only prompting (001) reaches 2.99\% WER, with SIM falling to 0.7555 (Table~\ref{tab:prompt_cv3}).
Speech-rate correlations remain close (0.832/0.829/0.827).
These results favor separating autoregressive planning from acoustic reference use; 001 remains conditioned through the FM prompt and speaker embeddings.

Together, CFG and prompt routing provide complementary inference-time controls for adapting synthesis to different requirements without retraining.
CFG adjusts the balance between pronunciation accuracy and expressiveness, as reflected in the trade-off between WER and reference speech-rate correlation.
Selective reference-audio injection allows the transfer of timbre, prosody and style to be adjusted while retaining explicit speaker-identity conditioning.
This separation of speaker identity from acoustic reference use supports flexible voice cloning across styles, languages and even dialects, rather than requiring all reference characteristics to be reproduced together.

\section{Conclusion}
\label{sec:conclusion}

We presented \model{}, an LLM-based TTS model that generates continuous speech latents and is trained end to end.
A Gaussian bottleneck with a computable information-capacity upper bound enables joint representation learning for reconstruction, recognition and generation.
The capacity sweep shows that a larger information budget improves reconstruction but does not necessarily improve zero-shot synthesis, highlighting the need to balance acoustic fidelity with autoregressive predictability.
Experiments show complementary benefits from feedback-based optimization and reinforcement learning: the former improves intelligibility, while the latter further improves English and hard-case intelligibility alongside speaker similarity.
CFG strength and prompt routing provide complementary inference-time controls over intelligibility, speaker similarity and acoustic-detail following without retraining.
Together, these results support capacity-controlled continuous representations and complementary feedback-based and reward-based optimization for high-quality, controllable TTS.




\clearpage
\phantomsection
\addcontentsline{toc}{section}{References}
\bibliographystyle{plainnat}
\bibliography{references}

\reportappendix
\section{Capacity of the Bottleneck}
\label{app:capacity}

Let $\mu_{t,d}$ be the $d$-th dimension of the normalized latent, with $\mathbb{E}[\mu_{t,d}^2]=P_d$ and $\frac{1}{D}\sum_d P_d = 1$ by Eq.~\eqref{eq:rms}.
The observation $z_{t,d}=\mu_{t,d}+\sigma\epsilon_{t,d}$ is an additive Gaussian channel.
For a fixed power $P_d$, the mutual information $I(\mu_{t,d};z_{t,d})$ is maximized by a Gaussian input and equals $\tfrac12\log_2(1+P_d/\sigma^2)$.
Summing over dimensions and applying Jensen's inequality,
\begin{equation*}
  \sum_{d=1}^{D}\tfrac12\log_2\!\Big(1+\frac{P_d}{\sigma^2}\Big)
  \le \frac{D}{2}\log_2\!\Big(1+\frac{1}{\sigma^2}\Big),
\end{equation*}
with equality when all dimensions have equal power.
For $D=384$ at 25\,Hz, $\sigma=0.5$ gives 446 bits per frame (11.1\,kbps) and $\sigma=0.2$ gives 902 bits per frame (22.6\,kbps).
For a discrete tokenizer with $K$ codebooks of size $V$ at frame rate $f$, the corresponding bound is $fK\log_2 V$ bits per second, reached under uniform code usage.


\end{document}